\documentclass[a4paper,10pt]{article}
\pdfoutput=1 

\usepackage{jcappub} 

\usepackage{orcidlink}

\title{\boldmath A topologically charged four-dimensional wormhole and the energy conditions}

\author{Faizuddin Ahmed\,\orcidlink{0000-0003-2196-9622}}

\affiliation{Department of Physics, University of Science \& Technology Meghalaya, 793101, India}

\emailAdd{faizuddinahmed15@gmail.com; faizuddin@ustm.ac.in}

\abstract{In this research work, our primary focus revolves around the examination of a specific category of traversable wormholes known as topologically charged generalized Schwarzschild-Simpson-Visser-type wormhole, $ds^2=-\Big(1-\frac{2\,M}{\sqrt{x^2+b^2}}\Big)\,dt^2+\Big(1-\frac{2\,M}{\sqrt{x^2+b^2}}\Big)^{-1}\,\Big(\frac{dx^2}{\alpha^2}\Big)+(x^2+a^2)\,(d\theta^2+\sin^2 \theta\,d\phi^2)$. This wormhole is uniquely defined by a pair of key parameters including global monopole charge. A noteworthy outcome of our investigation is the observation that the energy-momentum tensor associated with this wormhole complies with both the weak energy condition (WEC) and the null energy condition (NEC). Furthermore, incorporation of global monopole charge introduces a substantial influence on the curvature properties of wormhole space-time and various associated physical quantities derived from this geometry.}


\arxivnumber{2308.00012}

\begin{document}
\maketitle
\flushbottom

\section{Introduction}
\label{sec:intro}

The concept of traversable wormholes was groundbreaking when Morris and Thorne introduced it in 1988 \cite{MT,MT2}. Unlike previously considered wormholes, such as the Einstein-Rosen bridge \cite{ER} (an analysis of Einstein and Rosen's work can be related to Flamm's earlier work \cite{LF1}, with English translations available in Refs. \cite{LF2,LF3}), or Wheeler's microscopic charge-carrying wormholes \cite{JAW}, traversable wormholes are defined in a way that allows for the two-way travel of objects. Before Morris and Thorne's work on wormhole physics, classical wormholes were discovered by H. G. Ellis \cite{HGE}, and related self-consistent solutions were found by K. Bronnikov \cite{KAB}. Both are static and spherically symmetric solutions involving a free phantom scalar field. Additionally, a time-independent spherically symmetric solution by T. Kodama \cite{TK} was known in the literature. While the practical feasibility of creating or finding such wormholes remains doubtful, their study has nevertheless paved the way for highly productive research in modern times. 

Numerous wormhole models have been constructed and thoroughly investigated, significantly contributing to our understanding of these objects. Some of these models include Visser's wormhole models \cite{MV}, a self-consistent wormhole by E. G. Harris \cite{EGH}, Teo's rotating wormhole \cite{Teo}, static and dynamic plane-symmetric wormholes \cite{jpsl}, Morris-Thorne wormholes with a cosmological constant \cite{jpsl2}, a general class of spherically symmetric wormhole solutions \cite{AD}, wormholes with varying cosmological constants \cite{FR}, stationary and cylindrically symmetric rotating cylindrical wormholes without exotic matter but not asymptotically flat \cite{KB,KB2}, a family of exact solutions to the Einstein–Maxwell equations for rotating cylindrically symmetric wormholes \cite{KB3}, NUT traversable wormholes \cite{GC}, and static, spherically symmetric wormhole solutions with a minimally coupled scalar field \cite{KB4}. Furthermore, a significant amount of work has been devoted to building and studying wormhole solutions within the framework of modified gravity theories \cite{fsnl4}, as well as in other modified theories like Brans-Dicke theory \cite{uu1,uu2}, $f(R)$ gravity theory \cite{fsnl,ob,AD2,MS}, conformal Weyl gravity \cite{uu3}, Einstein-Gauss-Bonnet theory \cite{uu4,uu8,uu9}, Lovelock gravity theory \cite{uu5,uu6,uu10}, Rastall gravity theory \cite{uu7}, Einstein-Cartan gravity \cite{kb2,MRM,MRM2,MRM3}, and scalar-tensor theory \cite{RVK,RVK2,RS,uu11,MZ}. In higher-dimensional theories, wormhole solutions have been constructed in various settings, including five-dimensional theory \cite{gd}, six-dimensional vacuum space-time \cite{avm}, and $n$-dimensional theory \cite{TT}, as well as with compact dimensions \cite{AD3}. Additionally, various other space-time geometries have been explored, each offering unique extensions and variations of well-known black hole and wormhole solutions. In Ref. \cite{ASMV}, a novel space-time was introduced, representing a one-parameter extension of the Schwarzschild black hole. The outcome of this study revealed a family of metrics that smoothly interpolate between a Schwarzschild black hole, a black-bounce geometry, and a traversable wormhole. Another intriguing investigation in Ref. \cite{gg1} focused on a regular Fisner space-time. This space-time was coupled with a massless canonical scalar field, leading to the emergence of a traversable wormhole. In Ref. \cite{gg2}, an extension of the Reissner-Nordstr\"{o}m space-time was reported, known as the black-bounce solution. This extension showcased intriguing properties and  characteristics. Furthermore, in Ref. \cite{fsnl0}, a wide-ranging collection of globally regular black bounce space-times was introduced. These space-times offer a generalization of the original Schwarzschild-Visser (SV) model and have become a subject of significant interest in the field. These studies represent only a fraction of the diverse range of space-time models explored in the literature. The exciting discoveries and innovative solutions found in these works continue to deepen our understanding of the complex nature of black holes, wormholes, and other fascinating aspects of theoretical physics.

However, a fundamental concern surrounding wormhole space-times is that they often represent exact solutions of the field equations that violate one or more energy conditions \cite{SWH}. Consequently, the stress-energy tensor associated with these solutions may not correspond to realistic physical matter, though it is not inherently impossible. It is widely held that any physically sensible system should adhere to the energy conditions. Among the various energy conditions, the weak energy condition (WEC) holds particular prominence. It asserts that the energy-density must be non-negative. In mathematical terms, for the energy-momentum tensor $T^{\mu\nu}$, this condition is expressed as $T_{\mu\nu}\,U^{\mu}\,U^{\nu} \geq 0$, where $U^{\mu}$ is the time-like four-velocity vector satisfying the condition $U^{\mu}\,U_{\mu}=-1$. The null energy condition (NEC), the weakest of these conditions, states that for any null vector field $k^{\mu}$, it should hold that $T_{\mu\nu}\,k^{\mu}\,k^{\nu} \geq 0$, where $k^{\mu}\,k_{\mu}=0$. Additionally, another crucial condition in the context of wormhole models is the average null energy condition (ANEC), which stipulates that the integral of $T_{\mu\nu}\,k^{\mu}\,k^{\nu}$ with respect to the affine parameter $\lambda$ along a complete null geodesic $\gamma$ with the tangent vector $k^{\mu}$ must be positive. Notwithstanding the challenges posed by the violation of the weak energy condition, the null energy condition, or the average null energy condition, the exploration of wormholes has ignited significant scientific curiosity and continues to drive ongoing research in various related fields. The theoretical investigation of these exotic space-time geometries continues to enhance our understanding of general relativity and other fundamental aspects of physics.  

Remarkable progress has been made in the field of wormhole space-times, addressing the longstanding challenge of exotic matter. It has been recognized that vacuum solutions of the field equations naturally satisfy the energy conditions. Building on this research, a recent breakthrough was reported in Ref. \cite{FRK}, where a defect wormhole was discussed that does not require exotic matter and adheres to both the weak and null energy conditions. In Ref. \cite{FRK2}, a vacuum defect wormhole was constructed, representing a significant advancement in this area. Furthermore, in Ref. \cite{FRK3}, a novel type of traversable wormhole solution was proposed, eliminating the need for exotic matter, marking a significant step forward in making wormholes more feasible and accessible. In this direction, a Schwarzschild-Klinkhamer traversable wormhole was introduced in Ref. \cite{FA}, incorporating a cosmic string and global monopole while maintaining compliance with the energy conditions. This development highlights the potential interplay between different physical phenomena in the context of wormholes and cosmos. Additionally, Ref. \cite{FRK4} introduced a Schwarzschild defect wormhole, enriching our understanding of the various types of traversable wormholes. Extending this wormhole research into higher dimensions, a five-dimensional defect wormhole, an extension of vacuum-defect, was presented in Ref. \cite{FRK5}. This advancement opens up new possibilities for exploring wormholes in broader contexts.

Certain Grand Unified Theories (GUTs) have proposed the formation of topological defects during the phase transition in the early universe through a spontaneous symmetry-breaking mechanism, as reported in Refs. \cite{kk2,kk3,kk4,kk6}.  One specific example of these topological defects is the global monopole, a spherically symmetric entity arising from the self-coupling triplet of scalar fields $\phi^a$ ($a=1,2,3$). This triplet of scalar fields undergoes spontaneous breaking of the global $O(3)$ gauge symmetry, resulting in the formation of cosmic strings. Global Monopoles (GMs) are characterized by their spherically symmetric nature and possess distinct properties compared to other topological defects. In Ref. \cite{MBAV}, Barriola {\it et al.} presented an approximate solution of Einstein's equations for the metric in the region outside the core of GM, which forms due to global symmetry breaking. Such a monopole possesses Goldstone fields with energy density that decreases with distance only as $r^{-2}$ leading to a linearly divergent total energy at large distances. They calculated its effect on the propagation of null geodesics, demonstrating that light experiences angular deflection even when neglecting the mass term of the GM. This effect is noteworthy because massive particles in the limit of zero GM mass would not experience any Newtonian gravitational attraction. Neglecting the mass term, the monopole metric describes a space with a deficit solid angle. The area of a sphere with radius $r$ is not $4\,\pi\,r^2$ but rather $4\,\pi\,\alpha^2\,r^2$ with $\alpha^2=(1-8\,\pi\,G\,\eta^2)<1$, where $G$ is the gravitational constant and $\eta$ is the energy scale at which the symmetry is broken (also called the vacuum expectation energy) \cite{ERBM}. The values of the dimensionless quantity $8\,\pi\,G\,\eta^2 \simeq 10^{-6}$ and the vacuum expectation energy $\eta \sim 10^{16}$\mbox{Gev} \cite{MBAV}. When the GM mass term is not negligible, the solution describes a kind of Schwarzschild black hole with an additional GM charge, as discussed in \cite{nd}. Moreover, in Ref. \cite{Shi}, Shi {\it et al.} presented an exact solution to the non-linear equation describing the global monopole in flat space. They examined the metric outside GM and revealed a repulsive gravitational field outside the core, in addition to a solid angular deficit. It has been demonstrated in Ref. \cite{DPP} that the gravitational fields of GM can lead to the clustering of matter and Cosmic Microwave Background (CMB) anisotropies. Furthermore, GM may play a role in seeding the formation of supermassive black holes, as proposed in Reference \cite{RB}. Studies have indicated that this type of topological defect of GM induces a negative gravitational potential, resulting in a repulsive gravitational field \cite{Shi, kk12}. The same topic in asymptotically dS/AdS scenario was explored in Ref. \cite{nd2} and found that the gravitational potential can be either repulsive or attractive, depending on the value of the cosmological constant. Vilenkin {\it et al.} systematically expounded on the potential role of topological defects in our Universe in Ref. \cite{kk4}. Numerous researchers have illustrated the crucial role of topological defects in cosmological structure formation and the evolution of the cosmos, as referenced in \cite{MY,kk13,kk14,kk15,mm1,mm2,mm3,mm4,kk5,mm5,mm6}. Recently, there has been an investigation into wormholes in the Milky Way galaxy with a global monopole charge, as discussed in Ref. \cite{kk16}.

Our interest in the topological defects produced by GM charge has been motivated by a significant breakthrough uncovered by the MoEDAL detector, as detailed in Ref. \cite{BA}. The MoEDAL experiment is purposefully designed for the detection and exploration of magnetic monopoles and other highly ionizing, long-lived particles within the CERN Large Hadron Collider. The MoEDAL detector comprises three distinct detection systems: a nuclear track detector, a trapping detector array, and a time pixel array. In recent years, a detailed analysis of 13-TeV {\tt pp} collisions, utilizing the trapping detector between 2015 and 2017, has yielded crucial findings. Specifically, this analysis has established mass limits ranging from 1500 to 3750 GeV for magnetic charges of up to 5gD for monopoles possessing spins of 0, 1/2, and 1, as elaborated in Ref. \cite{BA2}. MoEDAL has also undertaken pioneering investigations by conducting the first-ever search for dyons. Through the utilization of a Drell–Yan production model, they have successfully excluded dyons with magnetic charges up to 5gD and electric charges up to 200e, accompanied by mass limits within the range of 870–3120 GeV. Additionally, MoEDAL has imposed constraints on monopoles bearing magnetic charges of up to 5gD, along with mass limits ranging from 870 to 2040 GeV, as detailed in Ref. \cite{BA3}.

Our aim in this research work is to explore the influence of topological charge produced by global monopoles or magnetic monopoles on the curvature properties of space-time and the matter-energy content in a two-parameter Schwarzschild-Simpson-Visser (SSV)-type wormhole. To accomplish this, we investigate this topologically charged two parameter wormhole space-time by solving Einstein's field equations and derive the expression of energy-density and pressures components, and notably, we discover that the energy-momentum tensor complies with both the weak and null energy conditions. One striking revelation is that the various physical quantities related to the space-time curvature remains finite, that is, regular everywhere including at the wormhole throat $x=0$, and tend to approach zero as $x$ approaches $\pm\,\infty$. This observation presents intriguing and potentially significant implications for the space-time's behavior in the vicinity of this wormhole configuration.

This paper centers on the study of a topologically charged Schwarzschild-Simpson-Visser (SSV)-type wormhole. Our investigation comprises three principal sections. In {\bf Section 2}, we introduce a metric ansatz for describing the topologically charged generalized SSV-type wormhole. We proceed to derive the corresponding field equations that govern the behavior of this metric. Moving on to {\bf Section 3}, we focus on a specific case of this generalized wormhole introduced earlier. Within this particular scenario, we thoroughly explore and discuss various properties associated with the wormhole. Finally, in {\bf Section 4}, we present our final discussions and conclusions based on the results obtained from the preceding two sections. In this paper, we employ a unit system where we set the speed of light, $c$, and the reduced Planck constant, $\hbar$, both equal to unity, that is, $c=1=\hbar$.

\section{A topologically charged generalized Schwarzschild-Simpson-Visser-type wormhole}

We extensively investigate a Schwarzschild-type wormhole solutions to the field equations, taking into account the influence of topological defects generated by global monopoles. We demonstrate that the presence of these topological defects alters the geometrical characteristics of the wormhole, and the matter-energy content adheres to both the weak energy condition and the null energy condition. 

Therefore, we would like to introduce a topologically charged wormhole space-time described by the following ansatz given by
\begin{eqnarray}
ds^2=-\Big(1-\frac{2\,M}{\sqrt{x^2+b^2}}\Big)\,dt^2+\Big(1-\frac{2\,M}{\sqrt{x^2+b^2}}\Big)^{-1}\,\frac{dx^2}{\alpha^2}+(x^2+a^2)\,(d\theta^2+\sin^2 \theta\,d\phi^2),
\label{1}
\end{eqnarray}
where $M$ represents mass of the objects, and two-parameter $(a, b)$ are non-zero positive constants obeying the condition 
\begin{eqnarray}
    0< a \leq b,\quad b>2\,M.\label{1a}
\end{eqnarray}
The different coordinates are in the ranges 
\begin{equation}
    -\infty < t < +\infty,\quad -\infty < x < +\infty,\quad \theta \in[0, \pi],\quad \phi \in (-\pi, \pi].\label{1b}
\end{equation}

In the limit $M \to 0$, the space-time described by the line-element (\ref{1}) simplifies into a topologically charged wormhole metric \cite{epjc}. Additionally, when $b$ equals $a$, and as $\alpha \to 1$, the same space-time (\ref{1}) reduces to the well-known Schwarzschild-Simpson-Visser wormhole \cite{ASMV}, which is a one-parameter modification of the Schwarzschild solution (as seen in Refs. \cite{gg1,gg2}). In our analysis, we are primarily focused on scenarios where $b \neq a$ and $\alpha \neq 1$. We will, however, consider the case where $b=a$ as a case in the next section. Therefore, the wormhole metric described by (\ref{1}) represents a two-parameter modification, with parameters $(a, b)$, of the Schwarzschild solution, and it's referred to as a topologically charged generalized Schwarzschild-Simpson-Visser-type wormhole. It's worth noting that when $b=0=a$, the metric (\ref{1}) yields a Schwarzschild-like solution with a global monopole.

Analysis of the radial null curves in this metric yields (setting $ds^2=0$, $d\theta=0=d\phi$):
\begin{equation}
    \frac{dx}{dt}=\pm\,\alpha\,\Big(1-\frac{2\,M}{\sqrt{x^2+b^2}}\Big).\label{1c}
\end{equation}
We can see that for $b > 2\,M$ and $x \in (-\infty, \infty)$, we have $\frac{dx}{dt} \neq 0$, so this geometry is in fact a (two-way) traversable wormhole in that condition.

It is worthwhile mentioning that we can define a “coordinate speed of light” for the space-time (\ref{1}) given by
\begin{equation}
    v(x)=\Big|\frac{dx}{dt}\Big|=\alpha\,\Big(1-\frac{2\,M}{\sqrt{x^2+b^2}}\Big).\label{1d}
\end{equation}
Therefore, an effective refractive index is given by
\begin{equation}
    n (x)=\frac{1}{\alpha}\,\frac{1}{\Big(1-\frac{2\,M}{\sqrt{x^2+b^2}}\Big)}.\label{1f}
\end{equation}

\subsection{Curvature tensor and its invariant }

In this part, we calculate the curvature tensor associated with the space-time geometry (\ref{1}) and discuss the effects of global monopole charge. Afterwards, we will calculate a few curvature invariants, such as the Ricci scalar $R$, the quadratic Ricci invariant $R_{\mu\nu}\,R^{\mu\nu}$, and the Kretschmann scalar $\mathcal{K}=R_{\mu\nu\rho\sigma}\,R^{\mu\nu\rho\sigma}$ and analyze them. 

The non-zero components of the curvature tensor $R^{\lambda}_{\mu\nu\sigma}$ for the space-time (\ref{1}) are given by
\begin{eqnarray}
    &&R^{t}_{xxt}=\frac{M\,(b^2-2\,x^2)}{(x^2+b^2)^{5/2}}\,\Big(1-\frac{2\,M}{\sqrt{x^2+b^2}}\Big)^{-1},\nonumber\\ &&R^{t}_{\theta\theta t}=\frac{M\,\alpha^2\,x^2}{(x^2+b^2)^{3/2}},\nonumber\\
    &&R^{t}_{\phi\phi t}=R^{t}_{\theta\theta t}\,\sin^2 \theta,\nonumber\\
    &&R^{x}_{txt}=\alpha^2\,\Big(1-\frac{2\,M}{\sqrt{x^2+b^2}}\Big)^2\,R^{t}_{xxt},\nonumber\\
    &&R^{x}_{\theta\theta x}=\frac{\alpha^2\Big[M\,x^4+\sqrt{x^2+b^2}\,(a^2\,b^2+x^2)\,\Big(1-\frac{2\,M}{\sqrt{x^2+b^2}}\Big)\Big]}{(x^2+a^2)\,(x^2+b^2)^{3/2}},\nonumber\\
    &&R^{x}_{\phi\phi x}=R^{x}_{\theta\theta x}\,\sin^2 \theta,\nonumber\\
    &&R^{\theta}_{t \theta t}=R^{\phi}_{t \phi t}=\frac{1}{x^2+a^2}\,\Big(1-\frac{2\,M}{\sqrt{x^2+b^2}}\Big)\,R^{t}_{\theta\theta t},\nonumber\\
    &&R^{\theta}_{x \theta x}=R^{\phi}_{x \phi x}=-\frac{1}{x^2+a^2}\,\Big[a^2+\frac{M\,x^2\,(x^2+a^2)}{(x^2+b^2)^{3/2}}\,\Big(1-\frac{2\,M}{\sqrt{x^2+b^2}}\Big)^{-1}\Big],\nonumber\\
    &&R^{\theta}_{\phi\phi\theta}=-R^{\phi}_{\theta\phi\theta}\,\sin^2 \theta,\nonumber\\ 
    &&R^{\phi}_{\theta\phi\theta}=1-\frac{x^2\,\alpha^2}{(x^2+a^2)}\,\Big(1-\frac{2\,M}{\sqrt{x^2+b^2}}\Big).
    \label{4}
\end{eqnarray}
The Ricci scalar $R=g_{\mu\nu}\,R^{\mu\nu}$ is given by
\begin{eqnarray}
    R&=&\frac{2}{(x^2+a^2)^2(x^2+b^2)^{5/2}}\Big[-a^4\,M\,(b^2-2\,x^2)\alpha^2-x^6\sqrt{b^2 + x^2}(-1+ \alpha^2)\nonumber\\
    &&+b^4\,x^2\Big(2\,M\,\alpha^2-\sqrt{b^2+x^2}(-1+\alpha^2)\Big)-b^2\,x^4\,\Big(M\,\alpha^2+2\,\sqrt{b^2 + x^2}(-1+\alpha^2)\Big)\nonumber\\
    &&+a^2\,\Big\{2\,b^2\,x^2\,\Big(M\,\alpha^2+ \sqrt{b^2 + x^2}(1 - 2\,\alpha^2)\Big)+b^4\,\Big(4\,M\,\alpha^2+\sqrt{b^2+x^2}(1-2\,\alpha^2)\Big)\nonumber\\
    &&+x^4\,\Big(4\,M\,\alpha^2+\sqrt{b^2+x^2}(1-2\,\alpha^2)\Big)\Big\}\Big].\label{3a}
\end{eqnarray}
The quadratic Ricci invariant $R_{\mu\nu}\,R^{\mu\nu}$ is given by
\begin{eqnarray}
    R_{\mu\nu}\,R^{\mu\nu}&=&\frac{1}{(x^2+a^2)^4\,(x^2+b^2)^5}\,\Bigg[M^2\,(a^2+x^2)^2\,\Big(3\,b^2\,x^2+a^2\,(b^2-2\,x^2)\Big)^2\,\alpha^4\nonumber\\
    &&+2\,(a^2+x^2)^2\,(x^2+b^2)^5\,\Big\{1+\Big(-1+\frac{2\,M\,b^2}{(x^2+b^2)^{3/2}}\Big)\,\alpha^2\Big\}^2\nonumber\\
    &&+\Big\{3\,b^2\,M\,x^4+a^4\,M\,(b^2-2\,x^2)+2\,a^2\,\Big(x^4 \big(-3\,M+\sqrt{b^2+x^2}\big)\nonumber\\
    &&+b^4(-2\,M+\sqrt{b^2+x^2})+2\,b^2\,x^2\,(-M+\sqrt{b^2 + x^2})\Big)\Big\}^2\,\alpha^4 \Bigg].\label{3b}
\end{eqnarray}
The Kretschmann scalar or scalar curvature $\mathcal{K}=R_{\mu\nu\rho\sigma}\,R^{\mu\nu\rho\sigma}$ is given by
\begin{eqnarray}
    \mathcal{K}&=&\frac{2}{(x^2+b^2)^6\,(x^2+a^2)^4}\,\Big[8\,M^2\,x^4\,(a^2+x^2)^2\,(b^2+x^2)^3\,\alpha^4\nonumber\\
    &&+4\,M^2\,(b^2-2\,x^2)^2\,(a^2+x^2)^4\,(b^2+x^2)\,\alpha^4\nonumber\\
    &&+4\,(b^2+x^2)^3\,\Big\{M\,x^2\,(a^2+x^2)+a^2\,(b^2+x^2)\,\Big(-2\,M+\sqrt{b^2+x^2}\Big)\Big\}^2\,\alpha^4\nonumber\\
    &&+4\,(b^2+x^2)^3\,\Big\{M\,x^4+a^2\,\Big\{b^2\,\Big(-2\,M+\sqrt{b^2+ x^2}\Big)+x^2\,(-M+\sqrt{b^2+x^2})\Big\}\Big\}^2\,\alpha^4\nonumber\\
    &&+2\,(b^2+x^2)^5\,\Big\{(a^2+x^2)\,\sqrt{b^2+x^2}-x^2\,\Big(-2\,M+\sqrt{b^2+x^2}\Big)\,\alpha^2\Big\}^2\nonumber\\
    &&+2\,(b^2+x^2)^5\Big\{a^2\,\sqrt{b^2+x^2}+x^2\,\Big(2\,M\,\alpha^2-\sqrt{b^2+x^2}(-1+\alpha^2)\Big)\Big\}^2\Big].\label{3c}
\end{eqnarray}

From the preceding analysis, it is evident that the presence of topological defects in the space-time under investigation has a notable impact on the curvature properties. Various physical entities associated with the space-time curvature tensor, such as the Ricci tensor, the quadratic Ricci invariant, and the Kretschmann scalar as well as the curvature tensor components are significantly influenced by the topological defect produced by a global monopole charge $\alpha$.

In Figure 1, we illustrate the behavior of the Ricci scalar with respect to the coordinate $x$ for various parameter values. On the left plot, we set $\alpha=1/2$ and $M=1=a$. On the right plot, we fix $b=2.5\,M$ and $M=1=a$. As shown in the plots, increasing the values of the parameter $b$ (where $b>2\,M$) and the global monopole parameter $\alpha$ leads to a notable decrease in the Ricci scalar. 

Figure 2 presents the behavior of the Kretschmann scalar as a function of the coordinate $x$ for various parameter values. On the left plot, we set $\alpha=1/2$ and $M=1=a$, while on the right plot, we use $b=2.5\,M$ and $M=1=a$. As depicted in the plots, increasing the value of the parameter $b$ (where $b>2\,M$) results in a notable decrease in the Kretschmann scalar. This trend indicates that the space-time curvature is influenced by variations in parameter $b$. On the other hand, the Kretschmann scalar increases with increasing values of the parameter $\alpha$.

Figure 3 illustrates behavior of the quadratic Ricci invariant as a function of the coordinate $x$ for different parameter values. On the left plot, we set $\alpha=1/2$ and $M=1=a$, while on the right plot, we use $b=2.5$ and $M=1=a$. As depicted in the plots, increasing the value of the parameter $b$ (where $b>2\,M$) and the global monopole parameter $\alpha$ leads to a significant decrease in the quadratic Ricci invariant. 

Overall, these graphical representations further emphasize the important role played by the global monopole parameter $\alpha$ and parameters $(a,b)$ in shaping the curvature characteristics of the space-time. By understanding how these parameters influence the Ricci scalar, the Kretschmann scalar, and the quadratic Ricci invariant, we gain valuable insights into the intricate properties of the considered wormhole metric and its physical implications.

\begin{figure}[tbp]
    \includegraphics[width=2.6in,height=2.0in]{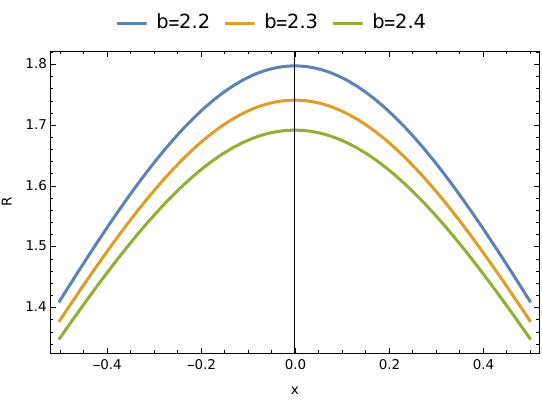}
    \hfill 
    \includegraphics[width=2.6in,height=2.0in]{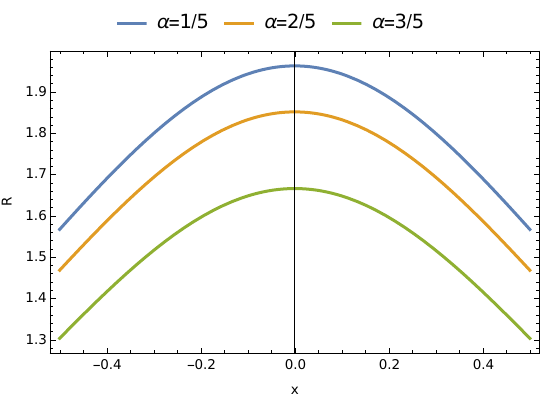}
    \caption{\label{fig: 1} The Ricci scalar}
    \hfill\\
    \includegraphics[width=2.6in,height=2.0in]{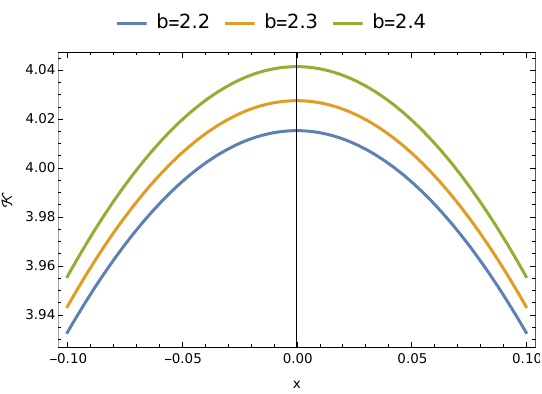}
    \hfill
    \includegraphics[width=2.6in,height=2.0in]{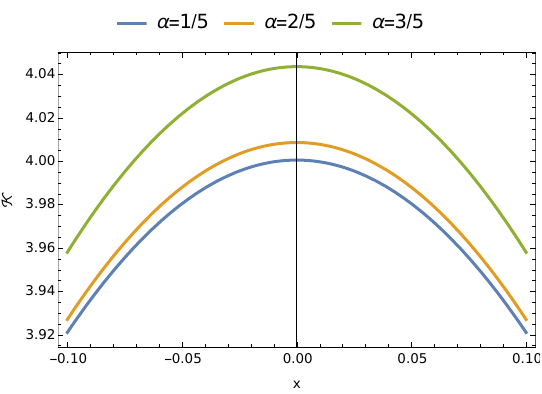}
    \caption{\label{fig: 2} The Kretschmann scalar}
    \hfill\\
    \includegraphics[width=2.6in,height=2.0in]{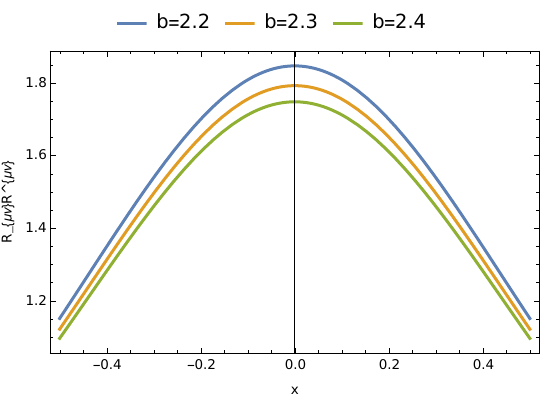}
    \hfill
    \includegraphics[width=2.6in,height=2.0in]{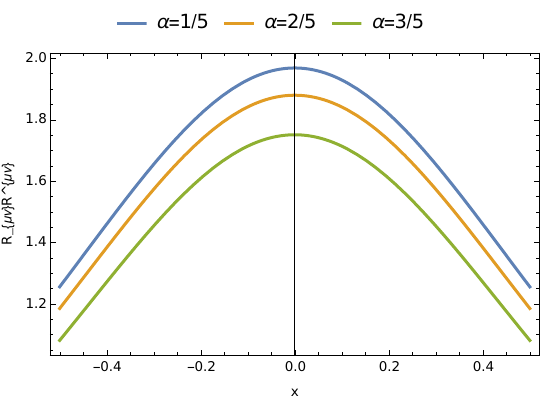}
    \caption{\label{fig: 3} The quadratic Ricci Invariant}
\end{figure}

One can easily find that at the wormhole throat, $x=0$, the different curvature invariants obtained in equations (\ref{3a})--(\ref{3c}) becomes  
\begin{eqnarray}
    R|_{x=0}&=&\frac{2}{a^2}\,\Bigg[1-\frac{(2\,b^3+a^2\,M-4\,b^2\,M)\,\alpha^2}{b^3}\Bigg],\nonumber\\
    R_{\mu\nu}\,R^{\mu\nu}|_{x=0}&=&\frac{1}{a^4\,b^6}\,\Big[a^4\,M^2\,\alpha^4+(2\,b^3+a^2\,M-4\,b^2\,M)^2\,\alpha^4+\frac{2}{b^6}\,\Big\{b+(-b+2\,M)\,\alpha^2\Big\}^2\Big],\nonumber\\
    \mathcal{K}|_{x=0}&=&\frac{4}{a^4\,b^6}\,\Big[M^2\,\alpha^4\,a^4+8\,b^4\,M\,(-b+M)\,\alpha^4+b^6\,(1+2\,\alpha^4) \Big]
    \label{5}
\end{eqnarray}
which are finite and vanishes for $x \to \pm\,\infty$. The non-zero components of the curvature tensor from equation (\ref{4}) at $x=0$ becomes
\begin{eqnarray}
    &&R^{t}_{xxt}=\frac{M}{b^3}\,\Big(1-\frac{2\,M}{b}\Big)^{-1},\nonumber\\
    &&R^{x}_{txt}=\frac{M\,\alpha^2}{b^3}\,\Big(1-\frac{2\,M}{b}\Big),\nonumber\\
    &&R^{x}_{\theta\theta x}=\alpha^2\,(b-2\,M),\nonumber\\
    &&R^{x}_{\phi\phi x}=R^{x}_{\theta\theta x}\,\sin^2 \theta,\nonumber\\
    &&R^{\theta}_{x \theta x}=R^{\phi}_{x \phi x}=-1=-R^{\phi}_{\theta\phi\theta},\nonumber\\
    &&R^{\theta}_{\phi\phi\theta}=-\sin^2 \theta
    \label{6}
\end{eqnarray}
which are finite with $b \neq 2\,M$.

\subsection{Matter-energy content and the energy conditions}

In this part, we will discuss the matter-energy distribution since the considered wormhole metric (\ref{1}) is a non-vacuum solution of the field equations. We analyze the physical quantities associated with this matter content and analyze the outcomes including the energy conditions. 

For space-time (\ref{1}), the nonzero components of the Ricci tensor $R_{\mu\nu}$ are given by
\begin{eqnarray}
    &&R_{tt}=\frac{M\,\alpha^2\,(3\,b^2\,x^2+a^2\,b^2-2\,a^2\,x^2)}{(x^2+b^2)^{5/2}\,(x^2+a^2)}\,\Big(1-\frac{2\,M}{\sqrt{x^2+b^2}}\Big),\nonumber\\
    &&R_{xx}=-\frac{M\,(b^2-2\,x^2)}{(x^2+b^2)^{5/2}}\,\Big(1-\frac{2\,M}{\sqrt{x^2+b^2}}\Big)^{-1}-\frac{2}{(x^2+a^2)^2}\times\nonumber\\
    &&\Bigg[a^2+\frac{M\,x^2\,(x^2+a^2)}{(x^2+b^2)^{3/2}}\Big(1-\frac{2\,M}{\sqrt{x^2+b^2}}\Big)^{-1}\Bigg],\nonumber\\
    &&R_{\theta\theta}=1-\alpha^2+\frac{2\,M\,\alpha^2\,b^2}{(x^2+b^2)^{3/2}},\nonumber\\
    &&R_{\phi\phi}=R_{\theta\theta}\,\sin^2 \theta.
    \label{2}
\end{eqnarray}

We choose a null vector $k^{\mu}$, time-like unit vector $U^{\mu}$, and spacelike unit vector $\eta^{\mu}$ along $x$-direction for the metric (\ref{1}) as follows: 
\begin{eqnarray}
    &&k^{\mu}=\Bigg[-\Big(1-\frac{2\,M}{\sqrt{x^2+b^2}}\Big)^{-1/2}, \alpha\,\Big(1-\frac{2\,M}{\sqrt{x^2+b^2}}\Big)^{1/2}, 0, 0\Bigg],\nonumber\\
    &&U^{\mu}=\Big(1-\frac{2\,M}{\sqrt{x^2+b^2}}\Big)^{-1/2}\,\delta^{\mu}_{t},\nonumber\\
    &&\eta^{\mu}=\alpha\,\Big(1-\frac{2\,M}{\sqrt{x^2+b^2}}\Big)^{1/2}\,\delta^{\mu}_{x},
    \label{7}
\end{eqnarray}
where these vector fields satisfies the following relations
\begin{eqnarray}
    -U^{\mu}\,U_{\mu}=1=\eta^{\mu}\,\eta_{\mu},\quad U_{\mu}\,\eta^{\mu}=0=k^{\mu}\,k_{\mu},\quad k_{\mu}\,\eta^{\mu}=1=U_{\mu}\,k^{\mu}.\label{77}
\end{eqnarray}

Let us examine the field equations $G_{\mu\nu}=\Big(R_{\mu\nu}-\frac{1}{2}\,g_{\mu\nu}\,R\Big)=T_{\mu\nu}$ by considering the following energy-momentum tensor form
\begin{eqnarray}
    T_{\mu\nu}=(\rho+p_{t})\,U_{\mu}\,U_{\nu}+p_{t}\,g_{\mu\nu}+(p_{x}-p_{t})\,\eta_{\mu}\,\eta_{\nu},
    \label{8}
\end{eqnarray}
where $\rho, p_x, p_t$ represents the energy-density, pressure component along $x$-direction, and tangential pressure, respectively. For space-time (\ref{1}), these physical quantities are given by
\begin{eqnarray}
    \rho=-T^{t}_{t}=-G^{t}_{t},\quad p_{x}=T^{x}_{x}=G^{x}_{x},\quad p_{t}=T^{\theta}_{\theta}=G^{\theta}_{\theta},
    \label{9}
\end{eqnarray}
where 
\begin{eqnarray}
    &&\rho=\frac{M\,\alpha^2\,(3\,b^2\,x^2+a^2\,b^2-2\,a^2\,x^2)}{(x^2+b^2)^{5/2}\,(x^2+a^2)}+\frac{R}{2}>0,\nonumber\\
    &&p_{x}=-\frac{M\,\alpha^2\,(b^2-2\,x^2)}{(x^2+b^2)^{5/2}}-\frac{2\,\alpha^2}{(x^2+a^2)^2}\,\Bigg[a^2\,\Big(1-\frac{2\,M}{\sqrt{x^2+b^2}}\Big)+\frac{M\,x^2\,(x^2+a^2)}{(x^2+b^2)^{3/2}}\Bigg]-\frac{R}{2},\nonumber\\
    &&p_{t}=\frac{1}{x^2+a^2}\,\Bigg[1-\alpha^2+\frac{2\,M\,\alpha^2\,b^2}{(x^2+b^2)^{3/2}}\Bigg]-\frac{R}{2},
    \label{99}
\end{eqnarray}
with $R$ given by equation (\ref{3a}).

Now, we check the validate of the different energy conditions. With the help of (\ref{7}), we see that 
\begin{equation}
    T_{\mu\nu}\,U^{\mu}\,U^{\nu}=\rho>0
    \label{10}
\end{equation}
a positive energy-density, indicating that the matter content satisfies the week energy condition (WEC) \cite{SWH}. Moreover, to satisfy the null energy condition (NEC), the relations $\rho+p_x>0$ and $\rho+p_{t}>0$ must holds good \cite{SWH}. In our case, we find 
\begin{eqnarray}
&&\rho+p_x=\frac{M\,\alpha^2\,(3\,b^2\,x^2+a^2\,b^2-2\,a^2\,x^2)}{(x^2+b^2)^{5/2}\,(x^2+a^2)}-\frac{M\,\alpha^2\,(b^2-2\,x^2)}{(x^2+b^2)^{5/2}}\nonumber\\
&&-\frac{2\,\alpha^2}{(x^2+a^2)^2}\,\Big[a^2\,\Big(1-\frac{2\,M}{\sqrt{x^2+b^2}}\Big)+\frac{M\,x^2\,(x^2+a^2)}{(x^2+b^2)^{3/2}}\Big],\nonumber\\
&&\rho+p_t=\frac{M\,\alpha^2\,(3\,b^2\,x^2+a^2\,b^2-2\,a^2\,x^2)}{(x^2+b^2)^{5/2}\,(x^2+a^2)}+\frac{1}{x^2+a^2}\,\Big[1-\alpha^2+\frac{2\,M\,\alpha^2\,b^2}{(x^2+b^2)^{3/2}}\Big].
\label{11}
\end{eqnarray}

It is worth noting that at the wormhole throat $x=0$, these physical quantities (\ref{99}) are finite given by
\begin{eqnarray}
    &&\rho|_{x=0}=\frac{1}{a^2}\,\Bigg[1-2\,\alpha^2\,\Big(1-\frac{2\,M}{b}\Big)\Bigg] > 0,\quad 0 < \alpha < 1,\nonumber\\
    &&p_{x}|_{x=0}=-\frac{1}{a^2},\nonumber\\
    &&p_{t}|_{x=0}=\frac{\alpha^2}{a^2}\,\Big(1-\frac{2\,M}{b}\Big)+\frac{M\,\alpha^2}{b^3}
    \label{12}
\end{eqnarray}
where we restricted $\alpha^2\,\Big(1-\frac{2\,M}{b}\Big) < 1/2$ with $b>2\,M$ since global monopole parameter $\alpha$ lies in the interval $0 < \alpha <1$ in gravitation and cosmology. These physical quantities associated with the energy-momentum tensor vanishes for $x \to \pm\,\infty$. From the above analysis, we observe that the null energy condition given by the relations $(\rho+p_x)<0$ everywhere even at $x=0$ and $(\rho+p_t)>0$. This indicates that even though the energy-density $\rho$ is positive everywhere, the null energy condition is partially satisfied. 

However, if one choose the global monopole parameter $\alpha$ is very small, that is, $\alpha <<1$, then the terms associated with $\alpha^2$ and its higher power should be neglected. In that situation, the various physical quantities obtained above approximate to as follows:
\begin{eqnarray}
    R &&\approx \frac{1}{(x^2+a^2)},\nonumber\\
    \mathcal{K} &&\approx \frac{1}{(x^2+a^2)^2},\nonumber\\
    R_{\mu\nu}\,R^{\mu\nu} &&\approx \frac{1}{(x^2+a^2)^2},\nonumber\\
    \rho &&\approx \frac{1}{(x^2+a^2)},\quad p_x \approx -\frac{1}{(x^2+a^2)},\quad p_{t} \approx 0
    \label{13}
\end{eqnarray}
which are finite at $x=0$ and vanishes for $x \to \pm\,\infty$.

Thus, from the above approximation values, we see that the matter-energy distribution satisfies both the weak energy condition and the null energy condition. Throughout the analysis, we notice that the presence of global monopole charge has significantly influenced all the physical properties associated with the space-time geometry and the physical quantities associated with the matter content, as a result, leads to notable changes in the results.

\section{A topologically charged Schwarzschild-Simpson-Visser-type wormhole}

In this section, we investigate a topologically charged Schwarzschild-Simpson-Visser (SSV)-type wormhole solution. This topologically charged SSV-type wormhole can be obtained by setting $b^2=a^2$ into the line-element described by (\ref{1}). Thereby setting $b^2=a^2$ into the metric (\ref{1}), we find a topologically charged SSV-type wormhole given by 
\begin{eqnarray}
      ds^2=-\Big(1-\frac{2\,M}{\sqrt{x^2+b^2}}\Big)\,dt^2+\Big(1-\frac{2\,M}{\sqrt{x^2+b^2}}\Big)^{-1}\,\Big(\frac{dx^2}{\alpha^2}\Big)+(x^2+b^2)\,(d\theta^2+\sin^2 \theta\,d\phi^2).
    \label{a1}
\end{eqnarray}

One can easily show that in the limit $\alpha \to 1$, this topologically charged wormhole (\ref{a1}) reduces to the well-known Schwarzschild-Simpson-Visser wormhole that has been discussed in details in Ref. \cite{ASMV}. In this work, we are mainly interested on analysing this particular topologically charged wormhole (\ref{a1}) and demonstrate that the presence of topological charge alters the geometrical characteristics of the metric, and influences physical quantities associated with the matter-energy distribution. We also show that the matter-energy content adheres to both the weak energy condition and the null energy condition.

\subsection{Curvature tensor and its invariant}

Here, we calculate the curvature tensor and then curvature invariants associated with the space-time (\ref{a1}) and discuss the outcomes. The Ricci scalar $R$ for the metric (\ref{a1}) is given by
\begin{eqnarray}
    &&R=\frac{2}{(b^2+x^2)}+\frac{6\,b^2\,M\,\alpha^2}{(b^2+x^2)^{5/2}}-\frac{2\,(2\,b^2+x^2)\,\alpha^2}{(b^2+x^2)^2}.\label{a3a}
\end{eqnarray}
The quadratic Ricci invariant $R_{\mu\nu}\,R^{\mu\nu}$ is given by
\begin{eqnarray}
    &&R_{\mu\nu}\,R^{\mu\nu}=\frac{1}{(b^2+x^2)^6\,\Big(-2\,M+\sqrt{b^2+x^2}\Big)^2}\,\Bigg[b^4\,M^2\,\Big(b^2+x^2-2\,M\,\sqrt{b^2+x^2}\Big)^2\,\alpha^4\nonumber\\
    &&+b^4\,\Big\{-7\,M\,x^2+6\,M^2\,\sqrt{b^2+x^2}+2\,x^2\,\sqrt{b^2+x^2}+b^2\,\Big(-7\,M+2\,\sqrt{b^2+x^2}\Big)\Big\}^2\,\alpha^4\nonumber\\
    &&+2\,(b^2+x^2)\,\Big(-2\,M+\sqrt{b^2+x^2}\Big)^2\,\Big\{(b^2+x^2)^{3/2}+\Big(2\,b^2\,M-(b^2+x^2)^{3/2}\Big)\,\alpha^2\Big\}^2 \Bigg].\label{a3b}
\end{eqnarray}
And the Kretschmann scalar curvatures is given by
\begin{eqnarray}
    &&\mathcal{K}=\frac{2\,\alpha^4}{(x^2+b^2)^5}\,\Big[4\,M^2\,x^4+2\,M^2\,(b^2-2\,x^2)^2+4\,\Big\{M\,x^2+b^2\,\Big(-2\,M+\sqrt{b^2+x^2}\Big)\Big\}^2\nonumber\\
    &&+2\,\Big\{\frac{(b^2+x^2)^{3/2}}{\alpha^2}+2\,M\,x^2-x^2\,\sqrt{b^2+x^2}\Big\}^2\Big].\label{a3c}
\end{eqnarray}

In Figure 4, we plot the Ricci scalar with $x$ and have shown that it is free from divergence at $x=0$. At the left one, we choose $\alpha=1/2$ and $M=1$. At the right one, we choose $b=2.3\,M$ and $M=1$. We see that by increasing the values of the parameter $b>2\,M$, and the global monopole parameter $\alpha$, the Ricci scalar gradually decreases and vanishes for $x \to \pm\,\infty$. 

In figure 5, we plot the Kretschmann scalar with $x$ and have shown free from divergence at $x=0$. At the left one, we choose $\alpha=1/2$ and $M=1$. At the right one, we choose $b=2.3\,M$ and $M=1$. By increasing the values of the parameter $b>2\,M$, the Kretschamnn scalar decreases and increases with increasing the values of $\alpha$.   

In figure 6, we plot the quadratic Ricci invariant with $x$ for different values of various parameters. At the left one, we choose $\alpha=1/2$ and $M=1$. At the right one, we choose $b=2.3\,M$ and $M=1$. By increasing the values of the parameter $b>2\,M$ and the global monopole parameter $\alpha$, the quadratic Ricci invariant decreases.

The non-zero components of the curvature tensor $R^{\lambda}_{\mu\nu\sigma}$ associated with the metric (\ref{a1}) are given by
\begin{eqnarray}
    &&R^{t}_{xxt}=\frac{M\,(b^2-2\,x^2)}{(x^2+b^2)^{5/2}}\,\Big(1-\frac{2\,M}{\sqrt{x^2+b^2}}\Big)^{-1},\nonumber\\ 
    &&R^{t}_{\theta\theta t}=\frac{M\,\alpha^2\,x^2}{(x^2+b^2)^{3/2}},\nonumber\\ 
    &&R^{t}_{\phi\phi t}=R^{t}_{\theta\theta t}\,\sin^2 \theta,\nonumber\\
    &&R^{x}_{txt}=\alpha^2\,\Big(1-\frac{2\,M}{\sqrt{x^2+b^2}}\Big)^2\,R^{t}_{xxt},\nonumber\\ 
    &&R^{x}_{\theta\theta x}=\frac{\alpha^2}{(x^2+b^2)^{3/2}}\,\Big[M\,x^2+b^2\,(-2\,M+\sqrt{x^2+b^2})\Bigg],\nonumber\\
    &&R^{x}_{\phi\phi x}=R^{x}_{\theta\theta x}\,\sin^2 \theta,\nonumber\\
    &&R^{\theta}_{t \theta t}=R^{\phi}_{t \phi t}=\frac{1}{x^2+b^2}\,\Big(1-\frac{2\,M}{\sqrt{x^2+b^2}}\Big)\,R^{t}_{\theta\theta t},\nonumber\\
    &&R^{\theta}_{x \theta x}=R^{\phi}_{x \phi x}=-\frac{1}{(x^2+b^2)^2}\,\Big[b^2+\frac{M\,x^2}{\sqrt{x^2+b^2}}\,\Big(1-\frac{2\,M}{\sqrt{x^2+b^2}}\Big)^{-1}\Big],\nonumber\\
    &&R^{\theta}_{\phi\phi\theta}=-R^{\phi}_{\theta\phi\theta}\,\sin^2 \theta\quad,\nonumber\\
    &&R^{\phi}_{\theta\phi\theta}=1-\frac{x^2\,\alpha^2}{(x^2+b^2)}\,\Big(1-\frac{2\,M}{\sqrt{x^2+b^2}}\Big).
    \label{a4}
\end{eqnarray}

\begin{figure}[tbp]
    \includegraphics[width=2.6in,height=2.0in]{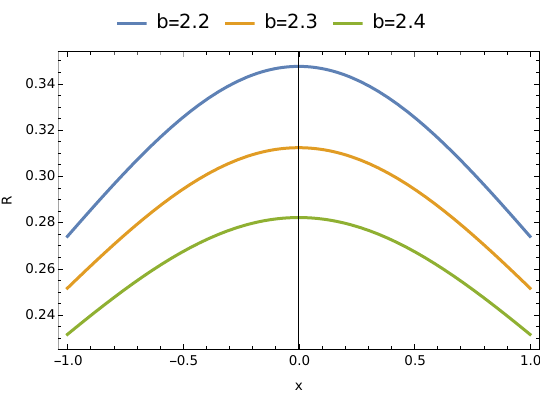}
    \hfill 
    \includegraphics[width=2.6in,height=2.0in]{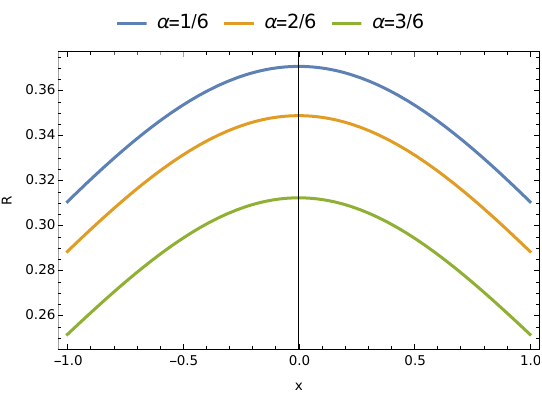}
    \caption{\label{fig: 4} The Ricci scalar}
    \hfill\\
    \includegraphics[width=2.6in,height=2.0in]{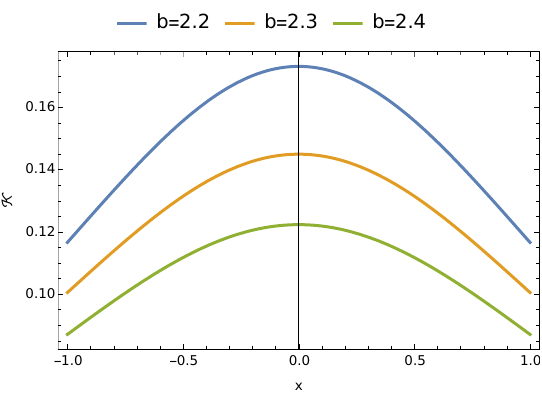}
    \hfill
    \includegraphics[width=2.6in,height=2.0in]{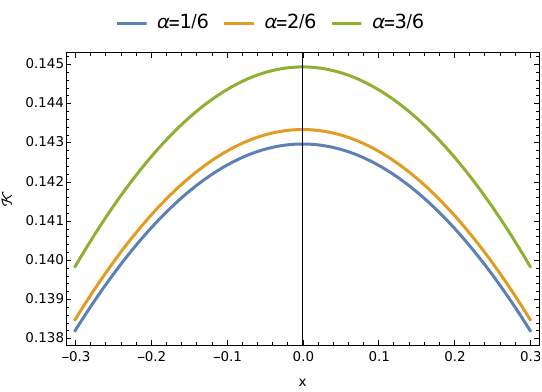}
    \caption{\label{fig: 5} The Kretschmann scalar}
    \hfill\\
    \includegraphics[width=2.6in,height=2.0in]{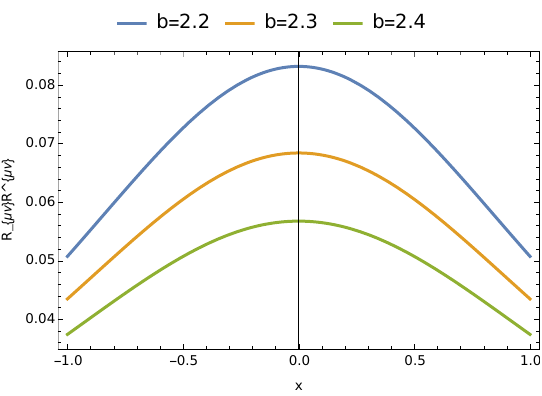}
    \hfill
    \includegraphics[width=2.6in,height=2.0in]{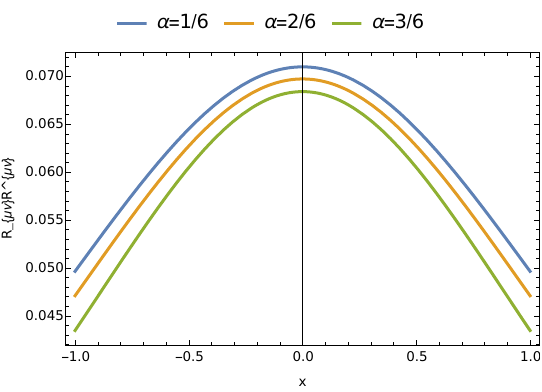}
    \caption{\label{fig: 6} The quadratic Ricci Invariant}
\end{figure}

At the wormhole throat $x=0$, the curvature tensor components (\ref{a4}) reduces to those components obtained in the equation (\ref{6}). Here, curvature invariants given in equation (\ref{a3a})--(\ref{a3c}) at $x=0$ are finite given by 
\begin{eqnarray}
    R|_{x=0}&=&\frac{6\,M\,\alpha^2}{b^3}+\frac{2-4\,\alpha^2}{b^2}\nonumber\\
    R_{\mu\nu}\,R^{\mu\nu}|_{x=0}&=&\frac{1}{b^6}\,\Big[8\,b\,M\,\alpha^2+2\,M\,(-10\,b+9\,M)\,\alpha^4+b^2\,\Big(2-4\,\alpha^2+6\,\alpha^4\Big)\Big],\nonumber\\
    \mathcal{K}|_{x=0}&=&\frac{4}{b^6}\,\Big[M\,(-8\,b+ 9\,M)\,\alpha^4+b^2\,(1+2\,\alpha^4)\Big].
    \label{a5}
\end{eqnarray} 
From the above analysis, it is evident that the presence of global monopole charge $\alpha$ in the space-time under investigation has a significant impact on the curvature properties.

\subsection{Matter-energy content and the energy conditions}

In this part, we discuss the matter-energy distribution for the wormhole metric (\ref{a1}) and analyze the physical quantities associated with this matter content including the energy conditions. The non-zero components of the Einstein tensor $G_{\mu\nu}$ for the metric (\ref{a1}) are given by
\begin{eqnarray}
    &&G_{tt}=\Bigg[\frac{4\,M\,\alpha^2\,b^2}{(x^2+b^2)^{5/2}}+\frac{x^2+b^2-\alpha^2\,(2\,b^2+x^2)}{(x^2+b^2)^2}\Bigg]\,\Big(1-\frac{2\,M}{\sqrt{x^2+b^2}}\Big),\nonumber\\
    &&G_{xx}=\frac{\Big(-b^2+x^2\,(-1+\alpha^2)\Big)}{\alpha^2\,(x^2+b^2)^2}\,\Big(1-\frac{2\,M}{\sqrt{x^2+b^2}}\Big)^{-1},\nonumber\\
    &&G_{\theta\theta}=\frac{\alpha^2\,b^2\,(-M+\sqrt{x^2+b^2})}{(x^2+b^2)^{3/2}},\nonumber\\ 
    &&G_{\phi\phi}=G_{\theta\theta}\,\sin^2 \theta.
    \label{a2}
\end{eqnarray}

Now, we study the energy-momentum distribution for this topologically charged SSV-type wormhole (\ref{a1}). As done in the previous section, we consider here the anisotropic fluid (\ref{8})--(\ref{9}) as the matter-energy content. Solving the field equations, we get the energy-density, and the pressure components given by
\begin{eqnarray}
    &&\rho=\frac{4\,M\,\alpha^2\,b^2}{(x^2+b^2)^{5/2}}+\frac{\Big(x^2+b^2-\alpha^2\,(2\,b^2+x^2)\Big)}{(x^2+b^2)^2}>0,\nonumber\\
    &&p_{x}=\frac{-b^2+x^2\,(-1+\alpha^2)}{(x^2+b^2)^2},\nonumber\\
    &&p_{t}=\frac{\alpha^2\,b^2\,(-M+\sqrt{x^2+b^2})}{(x^2+b^2)^{5/2}}.
    \label{a6}
\end{eqnarray}

We see that the topological defects of GM influences the different physical quantities associated with the matter content given in equation (\ref{a6}). Thus, the results obtained in this analysis gets modification due to the GM charge compared to the result presented in Ref. \cite{ASMV}.

For the energy-momentum distribution anisotropic fluid, the null energy condition is that both the relations $\rho+p_x>0$ and $\rho+p_t>0$ for $b >2\,M$ must holds good. In our cases here, we obtain these relations as follows :
\begin{eqnarray}
    &&\rho+p_x=-\frac{2\,\alpha^2\,b^2}{(x^2+b^2)^2}\,\Big(1-\frac{2\,M}{\sqrt{x^2+b^2}}\Big)<0,\nonumber\\
    &&\rho+p_t=\bigg(\frac{3\,M\,\alpha^2\,b^2}{(x^2+b^2)^{5/2}}+\frac{1-\alpha^2}{x^2+b^2}\bigg)>0.
    \label{a7}
\end{eqnarray}

At the wormhole throat $x=0$, the different physical quantities (\ref{a6}) associated with the matter-energy becomes
\begin{eqnarray}
    &&\rho|_{x=0}=\frac{1}{b^2}\,\Bigg[1-2\,\alpha^2\,\Big(1-\frac{2\,M}{b}\Big)\Bigg],\nonumber\\
    &&p_{x}|_{x=0}=-\frac{1}{b^2},\nonumber\\
    &&p_{t}|_{x=0}=\frac{\alpha^2}{b^2}\,\Big(1-\frac{M}{b}\Big).
    \label{a8}
\end{eqnarray}
From the above analysis, we observe that $(\rho+p_x)<0$ and $(\rho+p_t)>0$ everywhere at $x \geq 0$. This indicates that even though the energy-density $\rho$ of the matter content is positive everywhere at $x \geq 0$, the null energy condition is partially satisfied. 

It is worthwhile mentioning that for small values of $\alpha << 1$, various physical quantities associated with space-time curvature tensor, such as the Ricci scalar ($R$), the quadratic Ricci invariant ($R^{\mu\nu}\,R_{\mu\nu}$), the Kretschmann scalar ($\mathcal{K}$) obtained in equations (\ref{a3a})--(\ref{a3c}) as well as those physical quantities given in equation (\ref{a7}) approximates to as follows:
\begin{eqnarray}
    R &&\approx \frac{1}{(x^2+b^2)},\nonumber\\
    R_{\mu\nu}\,R^{\mu\nu} &&\approx \frac{1}{(x^2+b^2)^2},\nonumber\\
    \mathcal{K} &&\approx \frac{1}{(x^2+b^2)^2},\nonumber\\
    \rho &&\approx \frac{1}{x^2+b^2}>0,\quad p_x \approx -\frac{1}{x^2+b^2},\quad p_t \approx 0.
    \label{a9}
\end{eqnarray}
One can now see that the matter-energy distribution anisotropic fluid satisfies both the weak energy condition and the null energy condition. Throughout the analysis, we noticed that the presence of GM charge characterized by the parameter $\alpha$ significantly influences all the physical properties associated with the space-time curvature and the matter content, as a result, leads to notable changes in the results compared to those obtained in Ref. \cite{ASMV}.

\section{Conclusions}

In Ref. \cite{ASMV}, the authors presented a regular black hole solution. This solution serves as an extension of the Schwarzschild solution with an additional parameter denoted as ``a", thus giving rise to the Schwarzschild-Simpson-Visser (SSV) model. Notably, the authors demonstrated that this solution represents a wormhole model, provided that the parameter $a >2\,M$. In our current work ({\it Section 2}), we have introduced a topological defect wormhole space-time (\ref{1}), which extends the two-parameter Schwarzschild wormhole by introducing parameters $(a, b)$. We showed that our wormhole solution is also an extension of the Schwarzschild-Simpson-Visser (SSV) model, now incorporating a topological defect generated by GM charge. As a result, we referred to our wormhole solution (\ref{1}) as the generalized Schwarzschild-Simpson-Visser model with a GM charge. By employing this metric, we successfully addressed Einstein's field equations, enabling us to determine the energy-density and pressure components of the matter-energy distribution. Crucially, we have demonstrated that the matter-energy satisfies the weak energy condition, even at $x=0$, and partly satisfies the null energy condition. It is worth noting that for small values of GM charge parameter $\alpha$ (where $\alpha <<1$), we have shown that the wormhole continues to meet the energy conditions. Throughout our investigation, we considered the parameter $a \leq b$, with $b >2\,M$, and made sure that GM parameter $\alpha$ is not equal to 1, that is, $ \alpha \neq 1$. Furthermore, we computed various fundamental physical quantities associated with the space-time curvature, such as the Ricci scalar, the quadratic Ricci invariant, and the scalar curvature. Our calculations revealed that all of these quantities remain finite at $x=0$. This finding assures the regularity and validity of the space-time at this point. 

In {\it Section 3}, we delved into a particular scenario that corresponds to the case where $a=b$ in the above analyzed wormhole space-time (\ref{1}) featuring a GM charge. In this section, we performed detailed computations for various physical quantities linked to the space-time curvature, including the Ricci tensor. By solving the field equations and considering anisotropic fluid as the matter content, we were able to determine a range of quantities associated with the fluid. Importantly, our analysis revealed that the matter content satisfied the weak energy condition, owing to the positive energy density of the fluid. Moreover, it partly satisfied the null energy condition throughout the space-time, including at the point $x=0$. The findings in both Sections 2 and 3 collectively indicate that the presence of the topological defect arising from the GM charge represented by the parameter $\alpha$ significantly influences both the geometrical and physical properties associated with the wormhole.   

In our future research endeavors, we are set to explore the geodesic motion of test particles and the bending of photon rays within the gravitational field of this topologically charged wormhole. Our goal is to thoroughly examine and analyze the outcomes of these investigations. By doing so, we aim to gain deeper insights into the gravitational lensing phenomenon induced by topological charge, shedding new light on this fascinating aspect of the wormhole's behavior.

\end{document}